\documentclass[pre,aps,reprint,twoside,showpacs,amsmath,amssymb]{revtex4-1}

\usepackage{graphicx}

\usepackage[utf8]{inputenc}

\usepackage{hyperref}

\newcommand{\Eq}[1]{Eq.~(\ref{#1})}
\newcommand{\Fig}[1]{Fig.~\ref{#1}}

\newcommand{\av}[1]{\left<{#1}\right>}

\begin{document}

\title{Improving the efficiency of extended ensemble simulations:\\
  The accelerated weight histogram method}

\author{Jack Lidmar}
\email{jlidmar@kth.se}

\affiliation{%
Theoretical Physics,
KTH Royal Institute of Technology,
AlbaNova,
SE-106 91 Stockholm,
Sweden}

\date{\today}

\begin{abstract}
  We propose a method for efficient simulations in extended ensembles,
  useful, e.g., for the study of problems with complex energy
  landscapes and for free energy calculations. The main difficulty in
  such simulations is the estimation of the \textit{a priori} unknown
  weight parameters needed to produce flat histograms.  The method
  combines several complementary techniques, namely, a Gibbs sampler
  for the parameter moves, a reweighting procedure to optimize data
  use, and a Bayesian update allowing for systematic refinement of the
  free energy estimate.  In a certain limit the scheme reduces to the
  $1/t$ algorithm of B.E.\ Belardinelli and V.D.\ Pereyra [Phys.\ Rev.\ E
  \textbf{75}, 046701 (2007)].  The performance of the method is
  studied on the two-dimensional Ising model, where comparison with
  the exact free energy is possible, and on an Ising spin glass.
\end{abstract}


\maketitle


\section{Introduction}
\label{sec:intro}

The complex behavior of models with rough energy landscapes (such as
spin glasses, biopolymers, etc.) is an important but challenging
problem. In many situations progress is possible only using computer
simulations, but this too is a notoriously difficult problem.  In
order to efficiently sample the equilibrium distribution of such
models it is necessary to overcome the barriers separating different
metastable minima, a process which can be very slow if the temperature
is low.  A particularly fruitful strategy to enhance the sampling is
to enlarge the configuration space to include some well-chosen
parameter(s) in the model. In simulated
tempering~\cite{Marinari1992,*Lyubartsev1992}, e.g., the temperature
is promoted to a dynamical variable, whereby the system heats up and
cools down randomly and gets a good chance to explore the energy
landscape.  Such \emph{extended ensemble} or \emph{generalized
  ensemble} methods have gained much attention recently and are
routinely used in simulations of such diverse problems as spin
glasses, biomolecules, and problems in statistics.  The methods are
also highly useful for free energy calculations and for the estimation
of the probability of extreme events.
An attractive feature is that they can easily be incorporated into
existing simulation methods.  The downside is, however, that in order
to work properly they require fine tuning certain \emph{a priori}
unknown weights.
The weights must be tuned to ensure that each parameter value
(e.g. temperature) of the extended ensemble is visited equally often
on average.  They are simply related to the free energy at the given
parameter value, and are therefore a highly useful byproduct of the
simulation,
\emph{if they can be estimated efficiently using some scheme}.
While several such schemes have been
constructed~\cite{Marinari1992,*Lyubartsev1992,Wang2001} there is a
strong need for improvements.
In this paper we propose one such scheme with a number of distinct
advantages.

Section \ref{sec:intro} gives a brief background on extended ensembles
and discusses some shortcomings of existing methods.
Section \ref{sec:AWH} introduces an improved method, the accelerated
weight histogram method.
In Sec.~\ref{sec:Bench} the method is tested and benchmarked on two
model problems, the two-dimensional Ising model and a
three-dimensional Ising spin glass.

\subsection{Extended ensembles}

We consider a model described by a probability distribution
$\pi_\lambda(x)$, which depends on one or more parameters $\lambda$.
Typically we want to study the model for a whole range of parameter
values.
In an extended ensemble simulation, states are sampled according to a
joint distribution $P(x,\lambda)$, which we express, without loss of
generality, as
\begin{equation}								\label{eq:joint}
  P(x, m) = \frac 1 {\mathcal Z} e^{f_m - E_m(x)},
\end{equation}
where $x \in \mathcal X$ denotes the configuration of the system and
we assume a discrete set of preselected parameter values $\lambda_m
\in \mathcal M = \left\{ \lambda_1, \lambda_2, \ldots,
  \lambda_M\right\}$.
The weights $e^{f_m}$ introduced in \Eq{eq:joint} allow tuning the
marginal distribution $P(m)$ to approach any desired form.
We assume that we have a way of generating samples from the
conditional distribution
\begin{equation}								\label{eq:conditional}
  P(x|m) \equiv \pi_m(x) = e^{F_m - E_m(x)}
\end{equation}
at fixed parameter $\lambda_m$, using, e.g., Markov chain Monte Carlo
(MC) or molecular dynamics (MD) methods.
Generally this can be done without knowledge of the normalization
constants $e^{-F_m}$.
In physics applications \Eq{eq:conditional} is often the ordinary
canonical distribution $\sim e^{-\mathcal E/T}$, where we absorbed the temperature into the
energy in order to treat it on equal footing as any other parameter of
the system. Likewise, $F_m$ denotes the dimensionless free
energy~\footnote{A common example is the simulated tempering
  ensemble, where $\lambda_m = T_m$ and $F_m = \mathcal F(T_m)/T_m$,
  with $\mathcal F$ equal to the real free energy.}.
In Bayesian statistics problems \Eq{eq:conditional} is typically the
\textit{a posteriori} distribution for the model parameters and
possibly missing data given a set of observations.

The ordinary (MC or MD) moves are then complemented with transitions
in parameter space, which in most cases consist of a nearest neighbor
random walk.  The weights $e^{f_m}$ need to be adjusted to make the
marginal distribution
\begin{equation}								\label{eq:marginal}
  P(m) = \sum_x P(x,m) = \frac 1 {\mathcal Z} e^{f_m - F_m}
\end{equation}
of $m$ approximately flat~\footnote{We leave aside the question of
  whether a flat distribution is really optimal. This will typically
  depend on the particular problem at hand. 
  One reasonable possibility would be, e.g., to make it proportional to the
  correlation time at $\lambda_m$. 
  We just note that a slight
  modification of our formalism allows for any prescribed target
  distribution $\pi_m$ (or even simpler let the density of
  parameter values be nonuniform).}.
This requires $f_m \approx F_m$ where $F_m$ is the exact
(dimensionless) free energy at $\lambda_m$, unknown at the beginning
of the simulation.

Quite generally existing methods to estimate the weights $e^{f_m}$ can
be divided into two different classes, \emph{iterative} and
\emph{dynamic}.  In an iterative method the weights are produced in a
sequence of preliminary runs, each run giving a better estimate than
the old, until sufficient accuracy is reached.  On the other hand, in
a dynamic scheme the weights are being continuously updated during a
long simulation.
The dynamic schemes have potential for faster convergence, but since
the weights are constantly changing, detailed balance is violated and
the samples collected cannot therefore be safely used to estimate
average values of interest.  In the iterative scheme the weights are
fixed during each run, and only updated between the runs.


\subsection{The Wang-Landau and $1/t$ methods}

One particularly elegant dynamic scheme is the Wang-Landau
method~\cite{Wang2001}, originally developed for simulations in the
closely related multicanonical ensemble~\cite{Berg1992}.
In this ensemble the state space is not extended, but instead one
replaces the Boltzmann weights of the ordinary canonical ensemble with
a different one aimed at producing a flat histogram of some quantity
$\lambda(x)$, usually the energy.  From an algorithmic point of view
the main difference
is that the elementary moves $x \to x'$ also change the value of
$\lambda(x)$, e.g., the energy, whereas in the extended ensemble
method they can be performed at constant $\lambda$.  The latter allows
for more flexibility when choosing the parameter moves, something we
exploit below.
The Wang-Landau method is straightforward to adapt to extended
ensemble simulations (as demonstrated in Ref.~\onlinecite{Zhang2007}).  Each
time the system visits a particular parameter $m$, the corresponding
free energy parameter $f_m$ is decreased by a certain amount, $f_m
\gets f_m - \delta f$. A histogram of visited parameter values is
collected and $\delta f$ is reduced by a factor, $\delta f \gets
\delta f / 2$, when the histogram meets a certain flatness
criteria. Then the histogram is reset and the process starts over with
the reduced modification constant.  The scheme where $\delta f$ is
halved each iteration anticipates an exponential convergence of $f_m$
to its true value $F_m$.  Unfortunately, this is not the case.  Instead,
the error saturates at a level where reasonably flat histograms are
produced, but the free energy estimate no longer improves since
$\delta f$ becomes too
small~\cite{Belardinelli2007,*Belardinelli2007a,Zhou2008}.
It has been realized~\cite{Belardinelli2007} that the modification
factor should rather be decreased at a slow steady rate $\delta f \sim
1/t$, where $t$ is the Monte Carlo time, without regard to any
histograms, at least during the later stages of a simulation. The
resulting $1/t$ method turns out to perform very well, both in
multicanonical and extended ensemble simulations.

\subsection{Open issues}

Nevertheless, there is still plenty of room for improvements.
What is, for example, the most efficient way to move around in
parameter space?
How can the data collected during the simulations be used most
effectively to produce an estimate of the free energy needed for
uniform sampling?
How should the estimates from different iterations, perhaps run in
parallel, be combined in an optimal way?
How should the set of parameters $\mathcal M$ be chosen?  Most often
the parameter moves form a nearest neighbor random walk, and then the
choice of the spacing between adjacent values may be a critical
issue. Having too large gaps between adjacent values may lead to small
acceptance rates and therefore very slow dynamics along the parameter
axis.
Having too densely spaced parameter values, on the other hand, can
make the dynamics of the random walk  itself a limiting factor,
again slowing down the dynamics.

\section{The accelerated weight histogram method}
\label{sec:AWH}

In this paper we propose an iterative scheme --- the accelerated weight
histogram method (AWH) --- which combines several different
complementary techniques to give a very efficient method which
addresses the issues mentioned above.
First of all, we allow large parameter steps by the use of a Gibbs
sampler (a.k.a. heat bath algorithm).  This is combined with a
reweighting procedure which makes optimal use of the information
collected during the moves.  Together these make it possible to choose
a rather densely spaced set of parameters, without being limited by
slow diffusion.
The free energy parameters are updated based on a histogram of
\emph{weights} (rather than a histogram of visited parameter values)
combined with the information collected during previous iterations.

The parameter moves are carried out as follows.  In the simplest case
we allow transitions $m \to m'$ to \emph{any} new parameter value
$\lambda_{m'}$, with a probability given simply by the conditional
probability of $m'$ given the current configuration $x$
\begin{equation}							\label{eq:transition-prob}
w_{m'm}(x) = P(m'|x)
= \frac{e^{-E_{m'}(x) + f_{m'}}}
{\sum\limits_{k \in \mathcal M} e^{-E_{k}(x) + f_{k}}} .
\end{equation}
The transition probabilities just calculated are accumulated in a
histogram of weights
\begin{equation}							\label{eq:W}
  W_k \gets W_k + w_{km}(x), \quad \forall k .
\end{equation}
Further, they can be used for on-the-fly reweighting of sampled
observables
\begin{equation}							\label{eq:reweight}
\av {A}_k =\frac {\sum_t A(x_t,k) w_{km_t}(x_t)} {\sum_t w_{km_t}(x_t)} ,
\end{equation}
where $\{x_t,m_t\}$ denote the time series of visited configurations.
The averages $\av{A}_k$ at a particular value $\lambda_k$ thus get
contributions from a whole range of parameter values.
Note that the validity of \Eq{eq:reweight} does not depend on the
$f_m$ being converged.
This reweighting scheme is akin to the optimal multihistogram
reweighting technique of Ferrenberg and Swendsen~\cite{Ferrenberg1989}
(but with no need to solve a nonlinear equation system).

The update procedure continues in an iterative way.
During each iteration a certain number, say $N_I$, of samples are collected
and then the free energy parameters are updated as
$f_k \gets f_k + \Delta f_k , \forall k$,
with
\begin{equation}								\label{eq:df}
  \Delta f_k = - \ln \left( \frac {W_k M}{N} \right) ,
\end{equation}
where $N$ is the total number of samples collected so far.
The weight histogram is then updated to reflect this change
\begin{equation}								\label{eq:update-Wk}
  W_k \gets W_k e^{\Delta f_k} = N/M ,
\end{equation}
i.e., the total weight collected is distributed evenly among the $M$
parameter values, and the next iteration starts.  Note that the
identity $N = \sum_k W_k$ holds before and after the update.
The histogram is thus \emph{not} reset to zero after the iteration but
continues to grow.  This makes the updates \Eq{eq:df} become smaller
and smaller and allows for finer and finer details of the free energy
to be resolved.

Equations \eqref{eq:transition-prob} to \eqref{eq:update-Wk} form the
core of the algorithm, which can be summarized as follows:
\begin{enumerate}
\item \label{algo:start}
  Perform $N_x$ updates of the configurations $x$ at fixed parameter value $\lambda_m$.
\item \label{algo:move}
  Perform a parameter move $m \to m'$ using the Gibbs
  sampler,~\Eq{eq:transition-prob}.
\item \label{algo:end-of-loop}
  Update the weight histogram using \Eq{eq:W} and sample any
  observables of interest using \Eq{eq:reweight}.
\item
  Repeat steps \ref{algo:start}-\ref{algo:end-of-loop} until $N_I$ samples have been obtained.
\item \label{algo:end-of-iteration}
  Update the free energy parameters $f_m$ using~\Eq{eq:df} and the
  weight histogram using \Eq{eq:update-Wk}.
\item
  Start a new iteration from step 1 unless the desired accuracy has been reached.
\end{enumerate}
%
%
One possible concern is that step \ref{algo:move} of the algorithm
requires the computation of $M = |\mathcal M|$ different quantities,
which can become time consuming if the set $\mathcal M$ is large (as
can easily happen in the case of two- or higher-dimensional parameter
spaces).  In practice, $w_{m'm}(x)$ will be exponentially small except
for a range of $m'$ close to $m$.  If this is the case one may limit
the search for the new state to a neighborhood $\Lambda \subset
\mathcal M$ of $m$ by replacing step \ref{algo:move} with
\begin{enumerate}
\item[2']
  Choose a subset of parameter values $\Lambda$ with probability
  $P(\Lambda|m)$.
  Perform a parameter move $m \to m' \in \Lambda$ using the Gibbs
  sampler [\Eq{eq:transition-prob}, but with the sum restricted to $\Lambda$].
\end{enumerate}
Detailed balance is maintained if $P(\Lambda|m) = P(\Lambda|m')$ for
all $m,m' \in \Lambda$.  A simple choice (in the one-dimensional case)
is to select a range of parameter values as an interval $\Lambda =
\left\{m-L, \ldots,m-L+R\right\} \cap \mathcal M$, where $L$ is a
random uniformly distributed integer in $[0,R]$ and $R$ is a
predetermined range.  The generalization to higher-dimensional
parameter spaces is straightforward.


\subsection{Bootstrapping the simulation}
\label{sec:bootstrap}

Clearly the update \Eq{eq:df} requires an initial guess for $f_k$ and
a positive value of $W_k = W_\text{prior}$ at the start of the
simulation.  This latter value can be seen as a Bayesian prior of our initial
guess of $f_k$, which is later on updated as new data becomes
available.  If we have reason to believe that the starting estimate of
the free energy is good (e.g., because the free energy is expected to
have small variations), we can use a large $W_\text{prior}$.
In many applications, however, our initial guess is going to be poor
and we need some kind of bootstrap to get an acceptable prior.
We propose the following heuristic scheme:
Carry out the same steps in the simulation as above, but in addition
check, after each iteration has completed (after step
\ref{algo:end-of-iteration}), whether all parameter values have been
visited a certain fixed (usually small $\sim$ 1--10) number of times.
If not, reset the number of samples $N \gets M'$, where $M'$
is the number of parameters visited so far, and let $W_k \gets N/M = M'/M$.
In this way the weight histogram does not start to accumulate data until
$M'=M$ whereby the free energy parameters will get relatively large
updates at the initial stages. Also one should avoid sampling
observables during this initial stage.
Alternatively one may use free energy perturbation or a few
Wang-Landau iterations to get a reasonable initial estimate of $f_m$.
After this, the simulation may proceed with an initial prior $W_k =1$.

It is further recommended to make each iteration quite short,
consisting of only $N_I \sim$ 100--1000 parameter moves, during
this initial stage. (Later on it may be increased.)
It is also advisable to monitor the histogram $H_m$ of visited parameter values,
although it is not used directly to update the free energy. The robustness of
the algorithm can then be increased by restarting the simulation if
the histogram gets too skewed, e.g., if the minimum value
$H_\text{min}$ is less than a certain fraction of the mean.
This could be an indication that initial nonequilibrium transients
have distorted the distribution of the collected samples, which would
violate the main assumption of the algorithm, namely that the samples
collected during each iteration follow \Eq{eq:marginal}.  If this
happens one should reset the weight histogram and the effective number
of samples (e.g., $W_k \gets H_\text{min}$, $N \gets M H_\text{min}$
or perhaps even $W_k \gets 1$, $N \gets M$),
to allow the simulation to recover from that situation.


\subsection{Combining several simulations}

Often it is advantageous to run simulations in parallel to make
efficient use of computational resources.  The scheme introduced above
can easily be adapted to such situations.  Each computing node $(n)$
runs an independent simulation (consisting of $N_n$ samples) leading
to an estimate $f_m^{(n)}$ of the free energy parameters.  These may then
be combined into a best estimate $\bar F_m$
\begin{equation}								\label{eq:combine1}
  e^{-\bar F_m} = \mathcal N \sum_n N_n e^{-f_m^{(n)}}
    \mathcal Z^{(n)} ,
\end{equation}  
where $\mathcal Z^{(n)} = \sum_k e^{f_k^{(n)} - \bar F_k}$ and $\mathcal 
N$ is an unimportant normalization constant. This equation is
easily solved by iterating
\begin{equation}								\label{eq:combine} 
\bar F_m \gets \bar F_m  - \ln\left(
  \frac{M\sum_n N_n e^{\bar F_m-f_m^{(n)}} \mathcal Z^{(n)}}
  {\sum_{m,n} N_n e^{\bar F_m-f_m^{(n)}} \mathcal Z^{(n)}}
\right),
\end{equation}
starting from one of the $f_m^{(n)}$
(and this usually converges within 2--5 iterations).
This way of organizing the simulation also has the advantage that
statistical errors can be estimated using the standard jackknife
method~\cite{Berg2004a} applied to \Eq{eq:combine}.


\subsection{Relation to the $1/t$ method}

Many variations of the basic algorithm are possible, and may be
related to other methods.  For example, it reduces to the $1/t$ method
in the limit obtained by the following modifications:
 (1) Replace the Gibbs sampler by a simple nearest neighbor Metropolis
      step.
 (2) Replace the weight histogram $W_m$ by a simple histogram $H_m$ of
      visited $m$.
 (3) Update the free energy parameters after \emph{every} step.
Since the histogram after a visit to $m$ is $H_k = N/M + \delta_{km}$, the
free energy update becomes $\Delta f_k = -\ln(H_k M/(N+1)) = -\ln(1 +
\delta_{km} M/N) + \ln(1 + 1/N) \approx - \delta_{km} M/N + 1/N$, where
the approximation holds when $N \gg M$.  The last term represents a
constant shift of all $f_k$ and can be dropped.  The resulting update
rule is thus simply $f_m \gets f_m - M/N$, leaving all other $f_k$
unmodified.
This corresponds exactly to the $1/t$ method~\cite{Belardinelli2007}
discussed earlier, and provides a new perspective on and additional
justification for that update scheme.

\begin{figure*}
\includegraphics[width=1\linewidth]{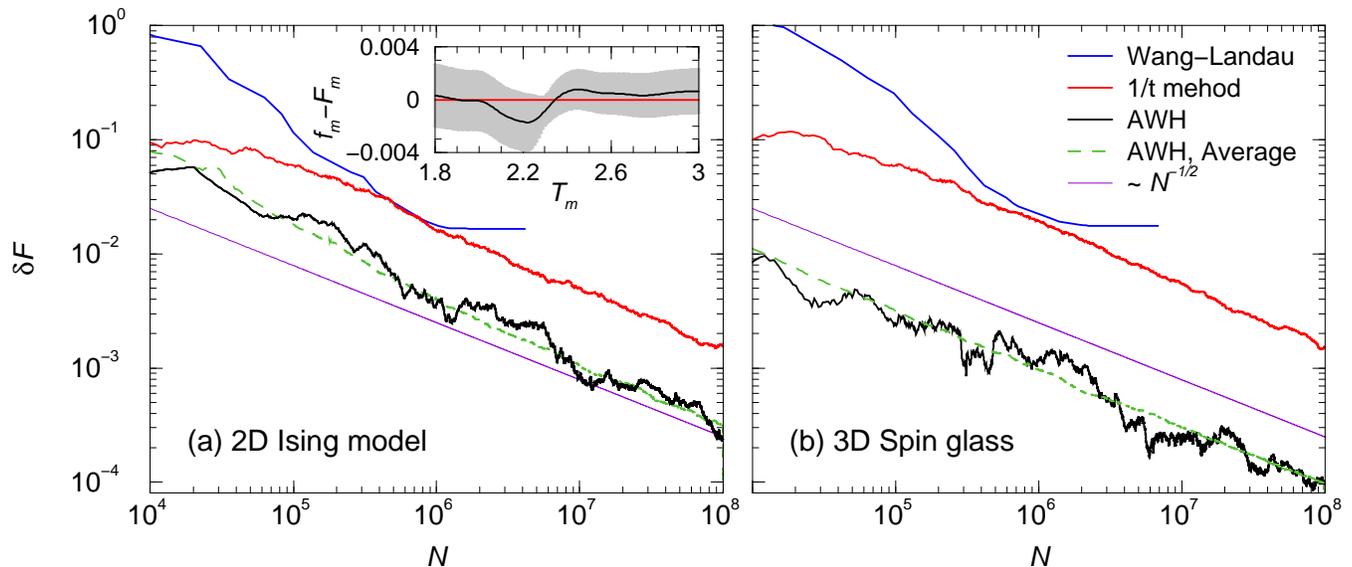}%
\caption{\label{fig:Ising}%
  (a)  Mean absolute deviation of estimated and exact free energy
  differences $\delta F$ of the $64 \times 64$ Ising model as a function of
  number of samples $N$.  From top to bottom: results from Wang-Landau
  iterations, the $1/t$ method, and the AWH method.
  Also included is the average behavior of the AWH method and a curve
  showing a $1/\sqrt N$ dependence. 
  Inset: Difference between estimated and exact free energy. The error
  bars (the shaded area) represent one standard deviation.
  (b) As in (a), but for an $8 \times 8 \times 8$ Ising spin glass.}
\end{figure*}


\section{Benchmarks of the method}
\label{sec:Bench}

To study the performance of the method and compare it with other ones
we apply it to the Ising model and a spin glass.  We carry out a
simulated tempering simulation, i.e., we choose as parameter $\lambda$
the temperature.  The algorithm alternates between ordinary canonical
Metropolis MC updates in which randomly chosen spins are flipped with
probability $\min(1,e^{-\beta \Delta \mathcal E})$, and updates which
change the temperature, leaving the spin configuration and the energy
$\mathcal E$ unchanged.  In the latter ones a new temperature $T_{m'}
= 1/\beta_{m'}$ is chosen with the probability
\begin{equation}								\label{eq:w-sim-T}
  w_{m'm}(\mathcal E) = \frac{e^{- \beta_{m'} \mathcal E + f_{m'}}}
  {\sum_k e^{-\beta_k \mathcal E + f_k}} .
\end{equation}

\subsection{Two-dimensional Ising model}

The two-dimensional (2D) Ising model is a common test case, since its
free energy can be calculated exactly~\cite{Ferdinand1969}.
We choose $M=128$ temperatures evenly spaced in the interval
$[1.8,3]$, which includes the critical temperature $T_c =
2/\ln(1+\sqrt 2) \approx 2.27$.  The system size is $L=64$ and we
use 100 000 iterations, each lasting for $1000$ MC sweeps, in total
$10^8$ sweeps, where each MC sweep corresponds to one update trial per
spin.  A temperature move is attempted after each MC sweep.

During the initial stages we use the scheme discussed in
Sec.~\ref{sec:bootstrap} to get an initial guess for the $f_m$ and a
prior weight $W_\text{prior}$:
At the start of the simulation we set $f_k = \beta_k \mathcal E_0$,
where $\mathcal E_0= -2L^2$ is the ground state energy, and $W_k=1/M$.
Then we check, after each iteration, whether all $M$ temperatures have
been visited at least twice during the simulation so far. If not, the
effective number of samples is reset to $N=M'/M$, where $M'$ is the
number of temperatures which actually were visited twice.
When all temperatures have been visited we have a reasonable initial
guess of $f_m$, and may continue the simulation as described in
Sec.~\ref{sec:AWH}, with $W_k \ge 1$.
Furthermore, we also monitor the histogram of visited temperatures to
look for anomalous deviations, which would indicate that the initial
guess was not so good after all.  Thus, we restart the simulation
(i.e., we set $N=M$, $W_k=1$ and reset the calculations of any
observables, but do not touch the $f_k$)
should the histogram of visited temperatures $H_m$ at some point fall
below $2\%$ of its mean.  This happened in about half of the
simulation runs, typically within the first 50 iterations.

To benchmark the method we plot, in \Fig{fig:Ising}(a), the mean
absolute deviation
\begin{equation}							\label{eq:mad}
  \delta F = \frac 1 {M-1} \sum_{m=1}^{M-1}
  \left| f_{m+1} - f_m - F_{m+1} + F_m \right|
\end{equation}
of consecutive free energy differences against the number of samples.
Here $F_m = \mathcal F(T_m)/T_m$ is the exact dimensionless free
energy.  For comparison we also include results from simulations using
Wang-Landau iterations (with flatness criteria $H_\text{min} > 0.4
H_\text{mean}$) and the $1/t$ method. For large times, the error for
both the $1/t$ and our method decrease as $1/\sqrt{N}$, whereas it
saturates for the Wang-Landau method.
For a given number of samples, the accuracy of the AWH method is almost
one order of magnitude better than the $1/t$ method.
The inset shows the difference between the final estimate, obtained by
combining 40 independent simulations using~\Eq{eq:combine}, and the
true free energy over the temperature range.  The error bars are
estimated using the jackknife method.

Another useful measure of the efficiency is the tunneling time, i.e.,
the time to go from the highest temperature to the lowest or
\textit{vice versa}.  This time was significantly reduced, nearly by a
factor of two, from $\sim$ 40 000 MC sweeps for the $1/t$ to
$\sim$ 21 000 for the AWH method.
It should be noted that the dynamics suffer severely from critical slowing down
in the vicinity of the phase transition, which constitutes a bottleneck
for the movement along the temperature axis.
While the extended temperature ensemble methods are effective for crossing
energy barriers, they do not overcome this slowing down by themselves.
In this sense the 2D Ising model (using single spin flip dynamics) is not a
particularly favorable test case.
However, the methods can easily be combined with cluster
methods, if available,
which do overcome the critical slowing down.
Replacing the single spin flip moves by, e.g., Wolff cluster
updates~\cite{Wolff1989} for $|T_m-T_c| < 0.1$ (the cluster moves being
most effective in the critical region) in the example above
practically eliminates the bottleneck and further reduces the
tunneling time by an additional factor $\approx 10$ to about 2200, for
the AWH method.  The $1/t$ method on the other hand only gained a
factor of two.

As discussed in Sec.~\ref{sec:AWH} one of the advantages of the AWH
method is the insensitivity to the spacing of parameter values
$\lambda_m$.
Indeed, varying the number of temperatures from $M = 32,
64, 128$ up to $256$, had negligible effect on the performance of the
algorithm, both in terms of the accuracy of the final free energy
estimate and the tunneling time, while the increase in the run time of
the simulation was marginal (and could be practically eliminated using
the update rule 2').
Upon decreasing $M$ below 16, on the other hand, the performance
quickly dropped.

\subsection{Three-dimensional Ising spin glass}

Next we apply the method to the three-dimensional Ising spin glass
with Gaussian couplings. This model has a disorder-dominated glass
phase at low temperatures $T < T_g \approx
0.95$~\cite{Katzgraber2006}, with a very rough energy landscape, making
it extremely challenging to study using conventional simulations.
The system size is $L=8$, and we use $M = 200$ temperatures logarithmically
spaced in $[0.7,3.5]$.
Figure \ref{fig:Ising}(b) compares the convergence of the different
methods for one particular random realization of the couplings.  As
there is no exact solution to compare with we use as reference instead
the best estimate obtained from 80 different runs (with an estimated
standard error $< 0.002$).
Here, the gain in accuracy, compared to the $1/t$ method, is more than
an order of magnitude.  The tunneling time, i.e., the time to go
between the high- and low-temperature extremes, is also significantly
shorter, by nearly a factor of $20$.

\section{Summary and Conclusions}

Let us reiterate the advantages of the AWH method:
Allowing for large steps gives a fast diffusion along the parameter
axis.  As a result, the spacing between neighboring values in the
discretized parameter space is not critical as long as it is small
enough and does not require any fine tuning to perform well.
We make efficient use of the data collected at all stages of the
simulation.  This is done by reweighting on the fly the samples taken
at the current parameter value to a whole range of different parameter
values. The information needed for this reweighting procedure is
essentially the same as what enables the large steps.
The data taken at earlier iterations are not thrown away, but are
instead used together with the new data to refine the estimate of the
free energy parameters.
Since the weights are constant during each iteration, the data
collected will, after an initial relaxation, be in equilibrium and
can be used for the calculation of any desired averages.

Altogether, these properties make up a very convenient method for
sampling models with rough energy landscapes, and for the
calculation of free energy differences.
It should be emphasized that it is the \emph{combination} of 
the Gibbs sampler, the reweighting scheme, and the update rule using
the weight histogram, which leads to the dramatic improvements.
The method is very general, is simple to implement, and can be applied to 
a broad range of problems in statistical physics, biophysics, 
statistics, etc.
Further improvements are likely, especially when it comes to the
heuristic scheme used during the early-stage bootstrap.


\acknowledgements

This work was supported by the Swedish Research Council (VR) through
Grant No.\ 621-2007-5138 and the Swedish National Infrastructure for
Computing (SNIC 001-10-155) via PDC.


\begin{thebibliography}{15}%
\makeatletter
\providecommand \@ifxundefined [1]{%
 \@ifx{#1\undefined}
}%
\providecommand \@ifnum [1]{%
 \ifnum #1\expandafter \@firstoftwo
 \else \expandafter \@secondoftwo
 \fi
}%
\providecommand \@ifx [1]{%
 \ifx #1\expandafter \@firstoftwo
 \else \expandafter \@secondoftwo
 \fi
}%
\providecommand \natexlab [1]{#1}%
\providecommand \enquote  [1]{``#1''}%
\providecommand \bibnamefont  [1]{#1}%
\providecommand \bibfnamefont [1]{#1}%
\providecommand \citenamefont [1]{#1}%
\providecommand \href@noop [0]{\@secondoftwo}%
\providecommand \href [0]{\begingroup \@sanitize@url \@href}%
\providecommand \@href[1]{\@@startlink{#1}\@@href}%
\providecommand \@@href[1]{\endgroup#1\@@endlink}%
\providecommand \@sanitize@url [0]{\catcode `\\12\catcode `\$12\catcode
  `\&12\catcode `\#12\catcode `\^12\catcode `\_12\catcode `\%12\relax}%
\providecommand \@@startlink[1]{}%
\providecommand \@@endlink[0]{}%
\providecommand \url  [0]{\begingroup\@sanitize@url \@url }%
\providecommand \@url [1]{\endgroup\@href {#1}{\urlprefix }}%
\providecommand \urlprefix  [0]{URL }%
\providecommand \Eprint [0]{\href }%
\providecommand \doibase [0]{http://dx.doi.org/}%
\providecommand \selectlanguage [0]{\@gobble}%
\providecommand \bibinfo  [0]{\@secondoftwo}%
\providecommand \bibfield  [0]{\@secondoftwo}%
\providecommand \translation [1]{[#1]}%
\providecommand \BibitemOpen [0]{}%
\providecommand \bibitemStop [0]{}%
\providecommand \bibitemNoStop [0]{.\EOS\space}%
\providecommand \EOS [0]{\spacefactor3000\relax}%
\providecommand \BibitemShut  [1]{\csname bibitem#1\endcsname}%
\let\auto@bib@innerbib\@empty
\bibitem [{\citenamefont {Marinari}\ and\ \citenamefont
  {Parisi}(1992)}]{Marinari1992}%
  \BibitemOpen
  \bibfield  {author} {\bibinfo {author} {\bibfnamefont {E.}~\bibnamefont
  {Marinari}}\ and\ \bibinfo {author} {\bibfnamefont {G.}~\bibnamefont
  {Parisi}},\ }\href {\doibase 10.1209/0295-5075/19/6/002} {\bibfield
  {journal} {\bibinfo  {journal} {Europhysics Letters}\ }\textbf {\bibinfo
  {volume} {19}},\ \bibinfo {pages} {451} (\bibinfo {year} {1992})}\BibitemShut
  {NoStop}%
\bibitem [{\citenamefont {Lyubartsev}\ \emph {et~al.}(1992)\citenamefont
  {Lyubartsev}, \citenamefont {Martsinovski}, \citenamefont {Shevkunov},\ and\
  \citenamefont {Vorontsov-Velyaminov}}]{Lyubartsev1992}%
  \BibitemOpen
  \bibfield  {author} {\bibinfo {author} {\bibfnamefont {A.~P.}\ \bibnamefont
  {Lyubartsev}}, \bibinfo {author} {\bibfnamefont {A.~A.}\ \bibnamefont
  {Martsinovski}}, \bibinfo {author} {\bibfnamefont {S.~V.}\ \bibnamefont
  {Shevkunov}}, \ and\ \bibinfo {author} {\bibfnamefont {P.~N.}\ \bibnamefont
  {Vorontsov-Velyaminov}},\ }\href {\doibase 10.1063/1.462133} {\bibfield
  {journal} {\bibinfo  {journal} {The Journal of Chemical Physics}\ }\textbf
  {\bibinfo {volume} {96}},\ \bibinfo {pages} {1776} (\bibinfo {year}
  {1992})}\BibitemShut {NoStop}%
\bibitem [{\citenamefont {Wang}\ and\ \citenamefont {Landau}(2001)}]{Wang2001}%
  \BibitemOpen
  \bibfield  {author} {\bibinfo {author} {\bibfnamefont {F.}~\bibnamefont
  {Wang}}\ and\ \bibinfo {author} {\bibfnamefont {D.~P.}\ \bibnamefont
  {Landau}},\ }\href {http://link.aps.org/doi/10.1103/PhysRevLett.86.2050}
  {\bibfield  {journal} {\bibinfo  {journal} {Physical Review Letters}\
  }\textbf {\bibinfo {volume} {86}},\ \bibinfo {pages} {2050} (\bibinfo {year}
  {2001})}\BibitemShut {NoStop}%
\bibitem [{Note1()}]{Note1}%
  \BibitemOpen
  \bibinfo {note} {A common example is the simulated tempering ensemble, where
  $\lambda _m = T_m$ and $F_m = \protect \mathcal F(T_m)/T_m$, with $\protect
  \mathcal F$ equal to the real free energy.}\BibitemShut {Stop}%
\bibitem [{Note2()}]{Note2}%
  \BibitemOpen
  \bibinfo {note} {We leave aside the question of whether a flat distribution
  is really optimal. This will typically depend on the particular problem at
  hand. One reasonable possibility would be, e.g., to make it proportional to
  the correlation time at $\lambda _m$. We just note that a slight modification
  of our formalism allows for any prescribed target distribution $\pi _m$ (or
  even simpler let the density of parameter values be nonuniform).}\BibitemShut
  {Stop}%
\bibitem [{\citenamefont {Berg}\ and\ \citenamefont
  {Neuhaus}(1992)}]{Berg1992}%
  \BibitemOpen
  \bibfield  {author} {\bibinfo {author} {\bibfnamefont {B.~A.}\ \bibnamefont
  {Berg}}\ and\ \bibinfo {author} {\bibfnamefont {T.}~\bibnamefont {Neuhaus}},\
  }\href {\doibase 10.1103/PhysRevLett.68.9} {\bibfield  {journal} {\bibinfo
  {journal} {Physical Review Letters}\ }\textbf {\bibinfo {volume} {68}},\
  \bibinfo {pages} {9} (\bibinfo {year} {1992})}\BibitemShut {NoStop}%
\bibitem [{\citenamefont {Zhang}\ and\ \citenamefont {Ma}(2007)}]{Zhang2007}%
  \BibitemOpen
  \bibfield  {author} {\bibinfo {author} {\bibfnamefont {C.}~\bibnamefont
  {Zhang}}\ and\ \bibinfo {author} {\bibfnamefont {J.}~\bibnamefont {Ma}},\
  }\href {\doibase 10.1103/PhysRevE.76.036708} {\bibfield  {journal} {\bibinfo
  {journal} {Physical Review E}\ }\textbf {\bibinfo {volume} {76}},\ \bibinfo
  {pages} {036708} (\bibinfo {year} {2007})}\BibitemShut {NoStop}%
\bibitem [{\citenamefont {Belardinelli}\ and\ \citenamefont
  {Pereyra}(2007{\natexlab{a}})}]{Belardinelli2007}%
  \BibitemOpen
  \bibfield  {author} {\bibinfo {author} {\bibfnamefont {R.~E.}\ \bibnamefont
  {Belardinelli}}\ and\ \bibinfo {author} {\bibfnamefont {V.~D.}\ \bibnamefont
  {Pereyra}},\ }\href {\doibase 10.1103/PhysRevE.75.046701} {\bibfield
  {journal} {\bibinfo  {journal} {Physical Review E}\ }\textbf {\bibinfo
  {volume} {75}},\ \bibinfo {pages} {046701} (\bibinfo {year}
  {2007}{\natexlab{a}})}\BibitemShut {NoStop}%
\bibitem [{\citenamefont {Belardinelli}\ and\ \citenamefont
  {Pereyra}(2007{\natexlab{b}})}]{Belardinelli2007a}%
  \BibitemOpen
  \bibfield  {author} {\bibinfo {author} {\bibfnamefont {R.}~\bibnamefont
  {Belardinelli}}\ and\ \bibinfo {author} {\bibfnamefont {V.}~\bibnamefont
  {Pereyra}},\ }\href {\doibase 10.1063/1.2803061} {\bibfield  {journal}
  {\bibinfo  {journal} {The Journal of chemical physics}\ }\textbf {\bibinfo
  {volume} {127}},\ \bibinfo {pages} {184105} (\bibinfo {year}
  {2007}{\natexlab{b}})}\BibitemShut {NoStop}%
\bibitem [{\citenamefont {Zhou}\ and\ \citenamefont {Su}(2008)}]{Zhou2008}%
  \BibitemOpen
  \bibfield  {author} {\bibinfo {author} {\bibfnamefont {C.}~\bibnamefont
  {Zhou}}\ and\ \bibinfo {author} {\bibfnamefont {J.}~\bibnamefont {Su}},\
  }\href {\doibase 10.1103/PhysRevE.78.046705} {\bibfield  {journal} {\bibinfo
  {journal} {Physical Review E}\ }\textbf {\bibinfo {volume} {78}},\ \bibinfo
  {pages} {046705} (\bibinfo {year} {2008})}\BibitemShut {NoStop}%
\bibitem [{\citenamefont {Ferrenberg}\ and\ \citenamefont
  {Swendsen}(1989)}]{Ferrenberg1989}%
  \BibitemOpen
  \bibfield  {author} {\bibinfo {author} {\bibfnamefont {A.~M.}\ \bibnamefont
  {Ferrenberg}}\ and\ \bibinfo {author} {\bibfnamefont {R.~H.}\ \bibnamefont
  {Swendsen}},\ }\href {\doibase 10.1103/PhysRevLett.63.1195} {\bibfield
  {journal} {\bibinfo  {journal} {Physical Review Letters}\ }\textbf {\bibinfo
  {volume} {63}},\ \bibinfo {pages} {1195} (\bibinfo {year}
  {1989})}\BibitemShut {NoStop}%
\bibitem [{\citenamefont {Berg}(2004)}]{Berg2004a}%
  \BibitemOpen
  \bibfield  {author} {\bibinfo {author} {\bibfnamefont {B.~A.}\ \bibnamefont
  {Berg}},\ }\href@noop {} {\emph {\bibinfo {title} {{Markov Chain Monte Carlo
  Simulations and Their Statistical Analysis}}}}\ (\bibinfo  {publisher} {World
  Scientific},\ \bibinfo {address} {Singapore},\ \bibinfo {year}
  {2004})\BibitemShut {NoStop}%
\bibitem [{\citenamefont {Ferdinand}\ and\ \citenamefont
  {Fisher}(1969)}]{Ferdinand1969}%
  \BibitemOpen
  \bibfield  {author} {\bibinfo {author} {\bibfnamefont {A.}~\bibnamefont
  {Ferdinand}}\ and\ \bibinfo {author} {\bibfnamefont {M.}~\bibnamefont
  {Fisher}},\ }\href {\doibase 10.1103/PhysRev.185.832} {\bibfield  {journal}
  {\bibinfo  {journal} {Physical Review}\ }\textbf {\bibinfo {volume} {185}},\
  \bibinfo {pages} {832} (\bibinfo {year} {1969})}\BibitemShut {NoStop}%
\bibitem [{\citenamefont {Wolff}(1989)}]{Wolff1989}%
  \BibitemOpen
  \bibfield  {author} {\bibinfo {author} {\bibfnamefont {U.}~\bibnamefont
  {Wolff}},\ }\href {\doibase 10.1103/PhysRevLett.62.361} {\bibfield  {journal}
  {\bibinfo  {journal} {Physical Review Letters}\ }\textbf {\bibinfo {volume}
  {62}},\ \bibinfo {pages} {361} (\bibinfo {year} {1989})}\BibitemShut
  {NoStop}%
\bibitem [{\citenamefont {Katzgraber}\ \emph {et~al.}(2006)\citenamefont
  {Katzgraber}, \citenamefont {K\"{o}rner},\ and\ \citenamefont
  {Young}}]{Katzgraber2006}%
  \BibitemOpen
  \bibfield  {author} {\bibinfo {author} {\bibfnamefont {H.~G.}\ \bibnamefont
  {Katzgraber}}, \bibinfo {author} {\bibfnamefont {M.}~\bibnamefont
  {K\"{o}rner}}, \ and\ \bibinfo {author} {\bibfnamefont {A.~P.}\ \bibnamefont
  {Young}},\ }\href {\doibase 10.1103/PhysRevB.73.224432} {\bibfield  {journal}
  {\bibinfo  {journal} {Physical Review B}\ }\textbf {\bibinfo {volume} {73}},\
  \bibinfo {pages} {224432} (\bibinfo {year} {2006})}\BibitemShut {NoStop}%
\end{thebibliography}
\end{document}